\documentclass [twocolumn,epsfig,prl,10pt,floatfix,amsmath,amssym]{revtex4}

\usepackage[dvips]{epsfig}

\usepackage{float}

\begin{document}

\title{A Geometric Theory of Diblock Copolymer Phases}
\author{Gregory M. Grason}
\author{B. A. DiDonna}
\author{Randall D. Kamien}
\affiliation{Department of Physics and Astronomy, University of Pennsylvania, Philadelphia, PA 19104-6396, USA}

\begin{abstract}
We analyze the energetics of sphere-like micellar phases in diblock copolymers in terms of
well-studied,  geometric quantities for their lattices.  We argue
that the A15 lattice with $Pm\bar{3}n$ symmetry should be favored as the blocks become more
symmetric and corroborate this through a self-consistent field theory.  Because phases with
columnar or bicontinuous topologies intervene, the A15 phase, though metastable, is not an equilibrium phase
of symmetric diblocks.  We investigate the phase diagram of branched diblocks and find that
the A15 phase is stable.
\end{abstract}
\date{\today}
\maketitle

\begin{figure}[b]
\center \epsfig{file=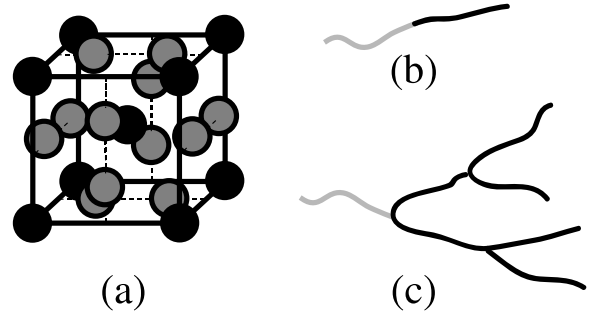,width=3in} \caption{In (a) we show the cubic
unit cell of the A15 lattice.  The black spheres form a BCC
lattice and the A15 lattice has, in addition, dimers on each face,
in grey.  The diblock topologies we consider are (b) linear and (c) branched with three generations.} \label{fig:a15}
\end{figure}

The ability to control the self-assembly of complex lattices by
manipulating molecular architecture remains an essential aspect in
the creation of new, functional materials.  With only a few
tunable parameters, diblock copolymer melts exhibit a wide variety
of equilibrium phases which can be understood via the mean-field
Gaussian chain model of ``AB'' diblock copolymers composed of
immiscible A and B blocks \cite{bates_sci1, mat_jphys}. Indeed, in
a system where the A and B-blocks are otherwise identical, there
are only two thermodynamic variables, $\phi$, the volume fraction
of $A$ type monomers, and $\chi N$, where $\chi$ is the
Flory-Huggins parameter characterizing the repulsive interactions
between the A- and B-type monomers and $N$ is the degree of
polymerization \cite{bates_arpc}. In this letter, we present a
model which predicts that the the A15 (shown in
Fig.~\ref{fig:a15}a) lattice of diblocks is stable relative to
other sphere-like phases for sufficiently large $\phi$ or, in
other words, sufficiently symmetric diblocks. We corroborate this
prediction by recalculating the phase diagram for symmetric
diblocks (Fig.~\ref{fig:a15}b) via a self-consistent field theory (SCFT) for diblock
copolymer melts \cite{mat_sch_prl}.

The ``classical'' diblock phases are well-understood:  near the
order-disorder transition (ODT), Leibler developed a Landau-like
theory in the weak-segregation regime to establish the stability
of a body-centered cubic (BCC) phase, a hexagonal phase of columns
and a lamellar phase \cite{leibler_macro}. Moreover, Semenov's
picture of spherical micelles interacting through a disordered
copolymer background when $\phi\ll 1$ accounts for the appearance
of the face-centered cubic (FCC) lattice near the ODT in the
mean-field phase diagram \cite{sem_macro}. The more exotic gyroid
phase was discovered \cite{hajduk_macro_94,
hajduk_macro_97a,hajduk_macro_97b} and was explained successfully
by Matsen and Schick via SCFT \cite{mat_sch_prl}.  In our study of
the A15 lattice, we find that the hexagonal and gyroid phases
intervene and thus there should be no stable A15 lattice for
simple diblocks.  However, sphere-like topologies are favored by
branched diblock copolymers \cite{olmsted_macro_98, gido_macro,
pickett} and dendritic polymers \cite{percec,percec2}; with this
in mind we predict that sphere-like phases are stabilized and that
the A15 phase is a ground state for this class of structures. By
implementing, to our knowledge, the first full SCFT treatment of
branched molecules (shown in Fig.~\ref{fig:a15}c) we have verified our theory.

In the dilute regime, Semenov's picture treats each micelle as an
undistorted sphere so that the outer block extends to a spherical
unit cell of radius $R_S$.  This unit-cell approximation provides
a lower bound for the free energy of the sphere-like phases
\cite{olmsted_macro_98}.  However, when the micelles assemble into
a periodic structure, the incompressibility of the melt allows no
interstitial gaps and the spheres must deform into the Voronoi or
Wigner-Seitz cells of the corresponding lattice. Our analysis will
be in the strong-segregation limit, away from the ODT where
fluctuation effects are less important. For small $\phi$, the
A-block will form the center of each sphere-like micelle.  As a
{\sl gedanken} experiment, we increase $\phi$ while maintaining
the topology and keeping the A-block on the inside.  In the most
extreme limit, $\phi\rightarrow 1$ and the A-blocks are surrounded
by a vanishingly thin B-block coat.
\begin{figure}
\center \epsfig{file=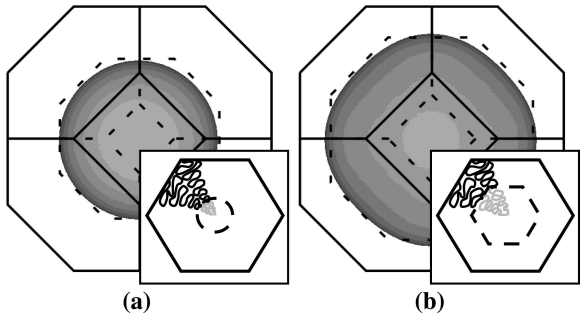, width=3.1in} \caption{A plot of the
AB interfaces in the BCC ($Im\bar{3}m$) phase obtained by
extracting the surface where the concentration of each block is
equal. Both figures were extracted for $\chi N=40$.  In (a)
$\phi=0.222$ and the interface is almost spherical.  In (b)
$\phi=0.45$ and the interface takes on the shape of the Voronoi
cell.  The solids lines represent the entire BCC cell, while the
dashed lines represent the reduced Voronoi cell that would arise
in the $\phi\rightarrow 1$ limit.  The insets depict (a) the
spherical interface limit (small $\phi$) and (b) the flat
interface limit (large $\phi$).} \label{fig: bcc}
\end{figure}

The free energy of strongly segregated diblock configurations of
spherical geometry arises from the tension in the AB interface and
the stretching of the polymers.  Though it may appear that the curvature of the interface must be considered, in neat systems the curvature energy is completely accounted for by stretching through the
incompressibility constraint \cite{mat_jphys}.  The energy of the AB interface is
proportional to its surface area $\Sigma$ (and $\sqrt{\chi}$ in
the strong segregation limit \cite{olmsted_prl_94,
olmsted_macro_98}). As illustrated in the insets to Fig.~\ref{fig:
bcc}, when $\phi$ grows, the A-blocks in the center of the micelle
fill a large volume and hence the curvature of the AB interface is
small.  As a result, the interface is easily deformed and we
expect that it will adopt the shape of the Voronoi cell.  A
dimensionless measure of the surface energy $\sqrt{\chi}\Sigma$ is
\begin{equation}
{\cal A} = \frac{\Sigma}{\Sigma_S}
\end{equation}
where $\Sigma_S$ is the area of the interface for a spherical
micelle of the same volume. This ratio may be rewritten as ${\cal
A}_X = \gamma(X)/(36\pi)^{1/3} $ where $\gamma(X)$ depends on
the lattice $X$ and relates the area to the
volume $V$ of the Voronoi cell: $\Sigma = \gamma(X) V^{2/3}$.  It is conjectured that
an A15 lattice of equal volume cells minimizes $\gamma$ \cite{weaire} and it is
known that $\gamma({\rm A15})=5.297<5.315=\gamma({\rm
BCC})<\gamma({\rm FCC}) = 5.345$ \cite{note1,jms}.  This fact can be
used to argue the stability of the A15 phase in fuzzy colloidal
systems \cite{kamien_prl} and is crucial to the analysis here. The
area per unit volume scales as $R_S^{-1}$ and so the surface
tension contributes ${\cal A}/R_S$ to the free energy density.

The second contribution to the free energy arises from the
stretching of the polymers to fill the cell.  Due to incompressibility,
the number of chains in each wedge of solid angle $d\Omega$
is proportional its volume, $\frac{1}{3}R^3(\Omega)d\Omega$, where
$R(\Omega)$ is the distance to the boundary.  Since the stretching
energy for both blocks of chains in this wedge is proportional to
$R^2(\Omega)$ \cite{olmsted_prl_94},
\begin{equation}\label{eq:I}
{\cal I}_X= \frac{\int_\Pi d{\Omega} \,R(\Omega)^5}{4\pi R_S^5}
\end{equation}
is a dimensionless measure of the stretching relative to the
stretching in a sphere of radius $R_S$ with the same volume as the Voronoi cell
$\Pi$.  This integral is proportional to the moment:
\begin{equation}
G(X) = \frac{\int_\Pi d^3\!R
\,R^2}{3\left(\int_\Pi d^3\!R\right)^{5/3}} = \frac{3^{2/3}}{5(4\pi)^{2/3}}{\cal
I}_X
\end{equation}
which has been well-studied \cite{Sloane} in the context of the
quantizing problem: $G({\rm BCC})=0.078543<0.078745=G({\rm
FCC})=G({\rm A15})$ \cite{moment}.   The stretching energy per
chain (or per unit volume) scales as $R_S^2$ and so the free energy contribution
from stretching is proportional to ${\cal I}R_S^2$.  The prefactor of this term will, in general,
depend on the topology of the micelles and $\phi$ \cite{olmsted_prl_94} but for the structures we consider, this prefactor does not change.

Putting these two effects together, we find a Flory-like free energy density:
\begin{equation}\label{eq:f}
f \sim  \frac{{\cal A}_X}{R_S} + {\cal I}_XR_S^2
\end{equation}
Minimizing over $R_S$, we find that $f=f_0 \left[{\cal I}_X{\cal
A}_X^2\right]^{1/3}\propto \gamma^{2/3} G^{1/3}$ where $f_0$ is
the energy of the spherical micelle \cite{olmsted_prl_94}.   It
follows, for instance, that as $\phi$ grows, the BCC lattice is
lower in energy than the FCC lattice since $\gamma({\rm BCC}) <
\gamma({\rm FCC}) $ \cite{olmsted_macro_98}.  We find that the A15
lattice is lower in energy still: $f_{A15} = 1.0707f_0 < 1.0722f_0
= f_{BCC} < f_{FCC} = 1.0772f_0$.  It should be noted that a more
detailed, self-consistent treatment \cite{olmsted_macro_98} yields
the same free energy (\ref{eq:f}) and the same numerical results
given here for the BCC and FCC lattices.  We have thus established
a fundamental and unavoidable geometric frustration between the
stretching and interfacial energies -- the BCC lattice minimizes
$G$ and the A15 lattice minimizes $\gamma$.   We note
that since the A15 lattice has two distinct sites, it is possible
for their corresponding cells to adopt different volumes to lower
the energy further.  Indeed, by adjusting the
Voronoi cells $G({\rm A15})$ and $\gamma({\rm A15})$ can be
lowered \cite{moment, jms}, ultimately lowering $f_{A15}$ by a
further 0.03\%.

As $\phi$ decreases the minority region shrinks and the curvature
grows, leading to a larger restoring force to the optimal,
spherical shape.  In this case the connectivity of the A-blocks to
the B-blocks would require a more detailed analysis, such as the
``kinked path'' {\sl ansatz} of Milner and Olmsted
\cite{olmsted_macro_98}.  Since the BCC lattice minimizes the
stretching energy it is not surprising that it is the equilibrium
phase at smaller $\phi$.
\begin{figure}
\epsfig{file=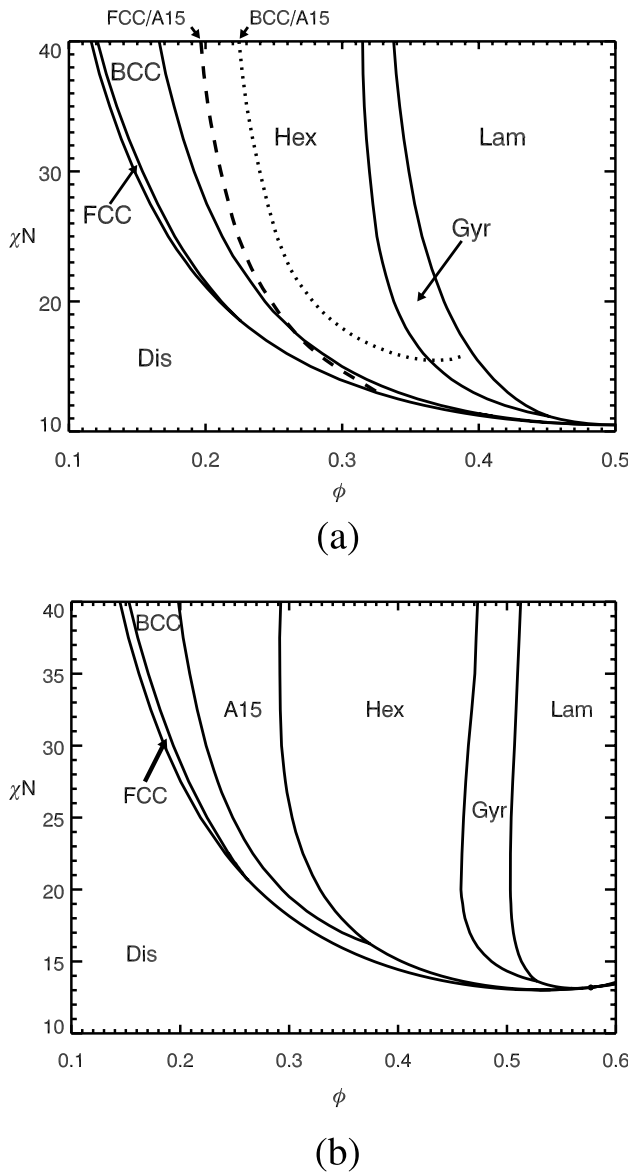,width=3.25in} \center \caption{The SCFT phase
diagram for melts composed of (a) linear chains. The dotted line is where the free energy of the BCC phase
is equal to that of the A15 phase, while the dashed line is the corresponding boundary for the FCC phase.  In the phase diagram (b) for the three-generation diblocks shown in Fig.~\ref{fig:a15}c the A15
phase is a ground state for a significant range of $\phi$ and
$\chi N$.  All the B blocks are of equal length.  Note the critical point on the lamellar/disordered boundary.}
\label{fig:3g}
\end{figure}

Because we assumed that the AB interface is perfectly flat, our
calculation puts upper limits on the free energies of these
sphere-like phases.  Nonetheless, our argument suggests that as
$\phi$ grows, the A15 phase should become the more stable
sphere-like phase. To critically test our theory, we employed the
SCFT implementation of Matsen and Schick to compare the free
energy of a phase with the $Pm\bar{3}n$ symmetry to the free
energy of the other sphere-like phases of diblock
melts~\cite{mat_sch_prl}. This scheme has the advantage of
providing numerically exact results using a Fourier basis of a
given symmetry and requires significantly less computing power
than is required to achieve an equivalent precision with a
numerical real-space approach \cite{fred_prl_99}. The efficiency
of this method allows us to explore the entire phase diagram and,
because we choose the symmetry of the basis we can quickly compare
three-dimensional lattices of different symmetries, even in
regimes where columnar, lamellar and gyroid phases are the ground
states \cite{mat_bates_macro_96}.  Note that when considering
close-packed lattices, hexagonal close-packed and FCC lattices
have degenerate free energies and so we use the $Fm\bar{3}m$ basis
only.   The first result of our calculation shows that as
$\phi\rightarrow 0$ the AB interface is spherical and as $\phi$
grows, the interface deforms significantly into the shape of the
Voronoi cell, as shown in Fig.~\ref{fig: bcc}.  This confirms our
intuitive discussion about the AB interface.

In order to compute the free energy of all phases within $0.005\%$
for $\chi N \leq 40$ we required up to 760 basis functions.  This
level of precision allowed us to delineate the phase boundaries to
within $\pm 0.001$ for $\phi$ and $\pm 0.01$ for $\chi N$.  Our
calculation corroborates our Flory-like theory and shows that the
A15 lattice is the most stable three-dimensional lattice for sufficiently large $\phi$,
as shown in Fig.~\ref{fig:3g}a.
In addition to the known phases for symmetric diblock copolymers,
we also computed the boundaries at which the free energies of the
three candidate sphere-like phases crossover.  We have reproduced
the boundary between the BCC and FCC phases reported by Matsen and
Bates \cite{mat_bates_macro_96}, and have also found the
boundaries at which the free energy of the A15 lattice becomes
lower than that of the of FCC and BCC phases, respectively.  Thus,
we see that the A15 lattice is the stable sphere-like phase as
$\phi\rightarrow 1$.  Note that the boundary between BCC and A15
terminates at $\phi = 0.390$ and $\chi N =15.86$ because at this
level of segregation for $\phi  >  0.390$ the $Im\bar{3}m$ sphere-like phase
melts.

Having considered the A15 lattice, it is natural to consider other
periodic arrangements of spheres.  We used the same SCFT method to
compute the free energy of sphere-like micelles assembled into the
simple cubic (SC) and diamond (D) lattice, with $Pm\bar{3}m$ and
$Fd\bar{3}m$ symmetries respectively.  For example, at $\chi N  =
17.0$ and $\phi = 0.26$ the free energies per chain in the BCC,
FCC, A15, SC and D phases are $3.2455$, $3.2459$, $3.2461$,
$3.2568$ and $3.2701$, respectively in units of $k_{B}T$.
This trend continues into the region where sphere-like phases are
not stable:  at $\chi N = 17.0$ and $\phi =
 0.31$ the free energies per chain of the A15, BCC, and FCC phases are
$3.6107$, $3.6120$, and $3.6158$,  respectively.  At this point of
the phase diagram the free energy per chain of the SC phase is
$3.6295$ and the D phase of spheres becomes unstable.  The free
energies of the BCC, FCC, and A15 phases remain within $1\%$ of
each other over the entire region of the phase space we explored
while the SC and D sphere-like phases are never the lowest energy
phases.

Though the A15 lattice will not be observed in neat linear diblock
melts, our calculation suggests that if the sphere-like phases
were stabilized it would persist as an equilibrium state.   One
way to stabilize sphere phases for more symmetric diblocks is to
change the architecture of the molecule. For instance, if one
block of the copolymers is branched, the stretching energy
combined with incompressibility will favor dividing surfaces which
are curved over a substantial portion of the phase diagram.   This
was predicted by Olmsted and Milner for miktoarm block copolymers
\cite{olmsted_macro_98} and later observed \cite{gido_macro}.
Similarly, calculations by Pickett for multiply branched diblocks
predict the same generic effect \cite{pickett}. We have
implemented a SCFT algorithm to explore the phase diagram of
branched diblocks with a linear A-block connected to $n$ B-blocks,
each of which are connected to $n$ B-blocks, {\sl etc.} for $g$
generations so that the final generation has $(g-1)n$ B-blocks
(the $g=1$ system is pure A).  We will report elsewhere on the
details of our calculation \cite{tobepub} but it closely follows
the original analysis of Matsen and Schick \cite{mat_sch_prl}.
Because of the time required and numerical intensity of the
calculation, we have focussed first on a $g=3$ generation diblock
with $n=2$ blocks at each branch as a ``proof-of-principle'' for
both our theoretical argument and our algorithm.  As shown in
Fig.~\ref{fig:3g}b, we find that sphere-like phases are stable over
a greater range of $\phi$ and that the A15 is a stable phase at
values of $\phi$ larger than those for which the BCC phase is
stable -- precisely what we would expect from our geometric
theory.  It is also worth noting that for this region of the phase
diagram, the gyroid is the equilibrium structure for this system,
and the double-diamond or hexagonally-perforated lamellar phases
are only metastable as in linear diblocks.

By employing purely geometric quantities we have argued that the
A15 lattice should be the lowest energy sphere-like micelle phase
of diblock copolymers as the blocks become more and more
symmetric.  Through a SCFT calculation we have corroborated our
theory though we found that where the A15 lattice was stable,
non-spherical columnar or gyroid phases were the ground
states of the diblock melt.  By considering branched diblocks
which favor sphere-like phases, we have found via SCFT that the
A15 is an equilibrium phase of branched diblocks with sufficiently
large (three) generations.  Further work will fully map out the phase
diagram of branched diblocks as function of molecular geometry and
topology.

\begin{acknowledgments}
It is a pleasure to thank V.~Percec, A.~Yodh and P.~Ziherl for
stimulating discussions.  We are indebted to D.~Duque, M.~Matsen
and M.~Schick for supplying us with their numerical code.  This
work was supported by NSF Grant DMR01-02459, the Donors of the
Petroleum Research Fund, Administered by the American Chemical
Society and a gift from L.J. Bernstein.
\end{acknowledgments}


\begin{thebibliography}{99}
\bibitem{note1}
Since the A15 lattice has two distinct Voronoi cells, the lattice quantities $\gamma(A15)$ and
$G(A15)$ represent the weighted average of the two cell types.
\end{thebibliography}
\end{document}